\documentstyle[preprint,aps,eqsecnum]{revtex}
\tighten
\begin{document}
\preprint{CMU-HEP-97-04; DOR-ER/40682-129}
\draft
\title{\bf GRAVITINO ZERO MODES ON $U(1)_R$ STRINGS}
\author{R.Holman, S.Prem Kumar}
\address{\it{Department of Physics, Carnegie Mellon University, \\ Pittsburgh, 
PA.15213, U.S.A.}}
\date{January 1997}
\maketitle
\begin{abstract}

We consider theories with a spontaneously broken {\em gauged} R-symmetry, which can
only occur in supergravity models. These give rise to cosmic R-strings upon
which gravitino zero modes can exist. We construct solutions to the
Rarita-Schwinger spin-3/2 equation describing the gravitino in the field of
these cosmic strings and show that under some conditions these solutions {\em may}
give rise to gravitino currents on the string. We discuss further mathematical
and physical questions associated with these solutions.
\end{abstract}
\pacs{11.27+d, 04.65.+e, 98.80.Cq}
\section{Introduction}

Supersymmetric models of particle physics interactions typically contain
operators that cause rapid proton decay which must be removed somehow if the
theory is to be phenomenologically consistent. One way of doing this is to
use R-parity\cite{rpar} as a symmetry of the Lagrangian. R-parity, if
unbroken, will also prevent the lightest supersymmetric particle (LSP) from
decaying, thus providing a candidate for WIMP dark matter\cite{griestetal}.

There are some potential problems with the use of R-parity, however. Though no
consistent theory of quantum gravity is yet known to exist, various conjectures
concerning ``low energy'' effects of gravitational fluctuations have been
made. In particular, the possibility that so-called wormhole configurations
might exist and have to be summed over in the Euclidean path integral
formulation of quantum gravity was at one point thought to be able to provide a
solution to the cosmological constant problem\cite{coleman}. While this
possibility was not realized\cite{nocosm}, it became clear that if wormhole
configurations did in fact have to be included in the functional integral for
gravity, then {\em global} symmetries would be in danger of being explicitly
broken by these effects. Essentially, global charge can ``pass'' through a
wormhole, going out of our universe. The sum over wormhole configurations then
induces an effective field theory in our universe that will necessarily include
operators that break the global symmetry explicitly.

An examination of these effects for R-parity\cite{gilbert} as well as for other
important global symmetries such as the Peccei-Quinn symmetry of axion
models\cite{usglobal,lindesuss} shows that wormholes can deliver devastating
blows to theories requiring such symmetries.

Wormholes can only destabilize {\em global} symmetries. Thus a natural solution
to the problem of how to keep R-parity as a good symmetry would be to gauge the
full $U(1)_R$ symmetry of which R-parity is a $Z_2$ subgroup. This can only be
done, however, in a {\em locally} supersymmetric theory due to the commutation
relations between the R generators and the supersymmetry generators; gauged R
symmetry requires gauged supersymmetry\cite{dzf}.

Recently, such gauged $U(1)_R$ models have been constructed for
phenomenological purposes\cite{dreiner,dzf2}. The $U(1)_R$ must be
spontaneously broken for consistency and as is well known, the breaking of a
$U(1)$ symmetry gives rise to cosmic string configurations\cite{vilshell}. What
is interesting about gauged R symmetry is that the gravitino must {\em
necessarily} carry a non-zero R charge. This then allows for the possibility
that gravitino zero modes may form on the string. This is what we investigate
in this work.

Fermion zero modes in topologically non-trivial backgrounds have been studied
extensively in the past, providing results that are interesting both from the
mathematical and physical points of view. The Dirac equation has been analysed
in various dimensions using kinks, vortices, 'tHooft-Polyakov monopoles and
instantons as background fields \cite{rebbi}. These studies have not only
yielded explicit solutions, but have also led to index theorems and to
connections with the axial anomaly.

Jackiw and Rossi \cite{jackiw} looked at Dirac fermions coupled to a charged
scalar in the background of an Abrikosov vortex in the $2$-D Abelian-Higgs
model. An index theorem for the above configuration was found by E. Weinberg
\cite{weinberg} and subsequently Witten \cite{witten} studied the Dirac fermion
in a cosmic string background in $(3+1)$-dimensions and found solutions
corresponding to transverse fermion zero modes trapped in the string, that gave
rise to superconducting currents. These latter results may have strong
astrophysical implications, insofar as the observational evidence for cosmic
strings is concerned\cite{ostrikercowie}.
     
In this article we will study spin-$3/2$ fermions in the background of a cosmic
R string in $(3+1)$ dimensions, using the Rarita-Schwinger equation. We look
for solutions to the gravitino field equations in such a string background,
considering two separate cases: (i) when $U(1)_R$ is broken but supergravity is
unbroken; (ii) both $U(1)_R$ and supergravity are broken. In the case where
supergravity is unbroken (i.e. the gravitino is massless) we find solutions
corresponding to fermions trapped on the string, with a non-zero fermionic
current along the string. We speculate that these solutions might also permit
superconducting currents. In the supergravity broken phase, we find solutions
that are localized on the string, and carry zero R-current. It must, however,
be emphasized, that we have found only {\em some} solutions for the system
considered; others may exist. We also present toy models of supergravity where
the $U(1)_R$ symmetry is spontaneously broken, while SUGRA (supergravity) is
broken in one 
case and left as an exact symmetry in another.

Section II is devoted to a brief review of R-symmetry and $N=1$ supergravity,
and the subsequent construction of 2 toy models. In Section III we obtain the
solutions to the gravitino field equations. Finally,
we present our conclusions and possible new lines of study in Section IV.

\section{R-invariance and supersymmetry}

We first present a brief review of $R$-symmetry in this section and then
proceed to construct simple toy models which incorporate this symmetry in a
consistent fashion. In global SUSY the $R$-transformations are defined in
terms of superfields as,
\begin{eqnarray}
\Phi(x,\theta,\bar{\theta})\hspace{0.1in}&{\rightarrow}&\hspace{0.1in}
e^{-in_{i}\alpha}\Phi(x,e^{-i\alpha} 
\theta,e^{i\alpha}\bar{\theta}),\label{rdef}\\
V(x,\theta,\bar{\theta})\hspace{0.1in}&{\rightarrow}&\hspace{0.1in}
V(x,e^{-i\alpha}\theta,e^{i\alpha}\bar{\theta}),\nonumber
\end{eqnarray}
where $\theta, \bar{\theta}$ are the Grassmann coordinates of superspace, $\Phi$
is a chiral superfield (scalar multiplet) with $R$-charge $n_{i}$ and $V$ is a
gauge vector multiplet. In addition, if we want the action to be
$R$-invariant, we must require that the superpotential $g$ have $R$-charge $=2$,
i.e.
\begin{equation}
g({\Phi}_{i})\hspace{0.1in}\rightarrow\hspace{0.1in} e^{-2i\alpha}
g({\Phi}_{i})
\end{equation}

In terms of the component fields, the $R$-transformation behaves like an axial
$U(1)$ transformation when acting on fermions. In particular,
fermions belonging to a scalar multiplet with charge $n_{i}$ will have an
axial charge $=n_{i}-1$, while all gauginos belonging to the gauge vector
multiplets will have an axial $R$-charge $1$. The scalar fields belonging to
the multiplet ${\Phi}_{i}$ have charge $n_{i}$. The above is true in general,
for both global and local $R$-transformations.

There is a subtlety associated with local $R$-symmetry, however. If we consider
the commutation relations between the R symmetry generator and the
supersymmetry generators we have\cite{baggerwess}:

\begin{equation}
\left[Q_\alpha, R\right]= i{(\gamma_5)_{\alpha}}^\beta\ Q_\beta.
\end{equation}
This implies that if the R-symmetry is gauged, this relation can only be
maintained if the $Q_\alpha$ are gauged as well, i.e. in a {\em locally}
supersymmetric theory\cite{dzf}. The important point to note is that, in the
locally $R$-symmetric theory, the gravitino and the $R$-photino both have axial
$R$-charges $1$:
\begin{eqnarray}
e^{\mu}_{m}\hspace{0.1in}&\rightarrow&\hspace{0.1in}e^{\mu}_{m}
\hspace{0.1in}\text{(graviton/vielbien fields)}\\\nonumber
\psi_{\mu}\hspace{0.1in}&\rightarrow&\hspace{0.1in}\exp(-i{\gamma}_{5}\alpha)
\psi_{\mu}\hspace{0.1in}\text{(gravitino)}\\\nonumber 
\lambda^{R}\hspace{0.1in}&\rightarrow&\hspace{0.1in}\exp(-i{\gamma}_{5}\alpha)
\lambda^{R}\hspace{0.1in}\text{($R$-photino)}\nonumber
\end{eqnarray}
\subsection{$N=1$ Supergravity}

An $N=1$ supergravity Lagrangian with canonical kinetic terms for the
Yang-Mills fields, is specified by the K$\ddot{\text{a}}$hler potential,
$G(\phi,\phi^*)$, which is a real, gauge-invariant function of the scalar
fields $\phi_{i}$. For our purposes, we will consider simple toy models, where
the matter multiplets are coupled {\em only} to the $U(1)_R$ gauge fields. In
such models, the scalar potential is given by,
\begin{eqnarray}  
-e^{-1}{\cal{L}}_P&=&\frac{1}{\kappa^4}
e^G[G^i(G^{-1})^{j}_{i}G_j-3]+\frac{1}{2\kappa^4}D^2
\\\nonumber
&\text{ }&\text{where}\\\nonumber
G^i&=&{\partial}G/{\partial}\phi_{i},\\\nonumber
G_i&=&{\partial}G/{\partial}\phi^{*i},\\\nonumber
G^i_j&=&{\partial}^2G/{\partial}\phi_{i}{\partial}\phi^{*j},\\\nonumber
\text{and}\hspace{0.1in}D&=&eG^in_{i}\phi_{i}.\nonumber
\end{eqnarray}
In addition, the function $G$ must be of the form,
\begin{eqnarray}
G(\phi,\phi^*)=J(\phi,\phi^*)+\ln|\kappa^3g(\phi)|^2
\end{eqnarray}
where $J$ is referred to as the K\"{a}hler metric and $g$ is the superpotential
which is an analytic (chiral) function of the fields. In this formulation of
supergravity \cite{ferrara} the gravitinos and gauginos have ``naive''
$R$-charge $=0$. However, to obtain the physical couplings one has to perform
the reverse chiral rotations on the gravitino field, as pointed out in
\cite{dreiner},
\begin{equation}
\psi_{{\mu}L}\hspace{0.1in}\rightarrow{\left(\frac{g^*}{g}\right)}^{1/4}
\psi_{{\mu}L} 
\hspace{0.4in}\psi_{{\mu}R}\hspace{0.1in}\rightarrow
{\left(\frac{g}{g^*}\right)}^{1/4}\psi_{{\mu}R}. 
\end{equation}
These rotations ensure that the gravitino has appropriate couplings to the
R-photon, and the gravitino sector of the Lagrangian is then given by,

\begin{eqnarray}
{\cal{L}}&=&-i{\bar{\psi}}_{{\mu}L}\gamma^{\mu\nu\lambda}[\partial_{\nu}
-ieA_{\nu}]{\psi}_{{\lambda}L}-
i{\bar{\psi}}_{{\mu}R}\gamma^{\mu\nu\lambda}[\partial_{\nu}+ieA_{\nu}]
{\psi}_{{\lambda}R}\label{gravitinolag}\\\nonumber
\\\nonumber
&+&\frac{1}{\kappa}e^{G/2}{\left(\frac{g}{g^*}\right)}^{1/2}{\bar{\psi}}_
{{\mu}L}\gamma^{\mu\nu}{\psi}_{{\nu}R} 
+\frac{1}{\kappa}e^{G/2}{\left(\frac{g^*}{g}\right)}^{1/2}{\bar{\psi}}_
{{\mu}R}\gamma^{\mu\nu}{\psi}_{{\nu}L}.\\\nonumber
\end{eqnarray}

The full $N=1$ lagrangian with Yang-Mills couplings to matter has a much more
complicated form and can be found in \cite{ferrara}. However, we are only
interested in {\em classical} solutions to the field equations obtained from
the supergravity lagrangian, and hence we restrict ourselves only to a
particular 
set of classical configurations for which the gaugino and other spin-$1/2$
fields in the theory vanish identically.  Therefore, quadratic terms which
generate gravitino-gaugino and gravitino-fermionic mixings can be ignored by
simply setting the gaugino and other fermionic fields to zero. Furthermore,
higher order, four-fermi type gravitino vertices are suppresed by inverse
powers of the Planck scale and hence are negligible.
 
Now any consistent gauge theory with axial vector couplings must be anomaly
free. Thus, if we set up a supergravity model with $U(1)_R$ symmetry, we must
ensure that the charges of the matter multiplets in the theory satisfy
appropriate anomaly cancellation conditions. Moreover, the superpotential must
have $R$-charge 2, and so every term in the superpotential must satisfy
${\Sigma}n_i=2$. The vanishing of the anomalous triangle diagram requires,

\begin{equation}
TrR^3=0.\label{trianom}
\end{equation}
Since gravity couples to all fields with equal strength, the mixed
gravitational anomaly can be removed by demanding,
\begin{equation}
TrR=0\label{gravianom}
\end{equation}
where the trace is over all the fermions in the theory.
\subsection{A Toy Model}
\subsubsection{Case I: SUGRA and $U(1)_R$ Broken}

With all the previously mentioned ideas in mind let us try to construct a toy
model where we can consistently implement the $R$-symmetry and break it
spontaneously along with supergravity. For simplicity we assume that $U(1)_R$
is the only gauge symmetry in the theory.  Let there be a set of chiral
multiplets $\{\Phi_0, \Phi_i\}$ ($i=1,2,...k$) with $R$-charges
$\{n_0,n_i\}$. We propose a non-canonical kinetic term for the scalar field
$\phi_0$ belonging to the multiplet $\Phi_0$. All other fields are assumed to
have canonical terms, so that the K\"{a}hler potential has the
following form,
\begin{eqnarray}
&G&(\phi,\phi^*)=-\frac{3}{2}\ln[(\phi_0+\phi^*_0)^2\kappa^2]+\Sigma\phi_i\phi
_i^* \kappa^2+\ln[\kappa^6|g(\phi_i)|^2]\\\nonumber
&g&(\phi_i)=m^2(\phi_1-\frac{{\sqrt{2}}\kappa^2}{3}\phi_1^2\phi_2)+\sum_i
\lambda_i\prod_j\phi_j^{L_{ij} }\\
&{L_{ij}}&n_j=2\label{superpot}\\
&m&\sim{\cal{O}}(1/\kappa).
\end{eqnarray}
Since the K\"{a}hler potential is gauge invariant, the above form
implies that $\phi_0$ must be neutral i.e. $n_0=0$.  The choice of the
superpotential (\ref{superpot}) fixes the $R$-charges of the fields $\phi_1$
and $\phi_2$ so that $n_1=2$ and $n_2=-2$. The choice of the non-canonical term
was of course, made arbitrarily. The only purpose of that term is to make our
calculations simple, by reducing the scalar potential into a sum of squares
whose minimum value is at zero. This ensures that the cosmological constant
vanishes. The scalar potential is then given by

\begin{eqnarray}
V(\phi,\phi^*)&=&\frac{1}{\kappa^4}e^{\kappa^2\Sigma|\phi_i|^2}
\frac{|g|^2\kappa}{|\phi_0+\phi_0^*|^3}\left[\sum_{i=1}^{k}\left|
\kappa^2\phi_i^* +\frac{g^i}{g}\right|^2\right]\\\nonumber
&+&\frac{1}{2\kappa^4}e^2\left(\sum_{i=1}^{k}n_i\phi_i\left[\kappa^2\phi_i^* 
+\frac{g^i}{g}\right]\right)^2. \nonumber
\end{eqnarray}
The charges of three of the multiplets were fixed uniquely by the form of the
superpotential. Additional multiplets with non-trivial $R$-charges will,
however, be required for purposes of anomaly cancellation. We denote the
charges of the fermions in the multiplets by $f_i\equiv n_i-1$. Therefore, $f_0=-1$,
$f_1=1$ and $f_2=-3$. The mixed gravitational anomaly and the triangle anomaly
are removed if,
\begin{eqnarray}
TrR&=&\sum_{i=3}^kf_i-1+1-3+1-21=\sum_{i=3}^kf_i-23=0\\\nonumber
TrR^3&=&\sum_{i=3}^kf_i^3+(-1)^3+1^3+(-3)^3+1^3+3\times1^3=
\sum_{i=3}^kf_i^3-23=0.\\\nonumber
\end{eqnarray}
The gravitino contribution to the mixed anomaly is $-21$ times, while the
contribution to the triangle anomaly is $3$ times that of the gaugino
\cite{dreiner}, \cite{gaume}.  The simplest solution is to
have $k=25$ i.e. introduce $23$ additional scalar multiplets each with
$R$-charge $=2$ into the superpotential in such a way that (\ref{superpot}) is
always satisfied. Thus, we can build a superpotential that fixes the charges of
the fields in such a way that the anomaly cancellation conditions are
satisfied:

\begin{equation}
g(\phi_i)=m^2(\phi_1-\frac{{\sqrt{2}}\kappa^2}{3}\phi_1^2\phi_2)+
\sum_{i=3}^{25} 
\lambda_i\phi_i^2\phi_2.
\end{equation}
There are other, more economical possibilities, but this one will suffice to
make our point. The field configuration that minimises the scalar potential
is,
\begin{eqnarray}
|\langle\phi_{1}\rangle|=
\frac{1}{\kappa},\hspace{0.2in}|\langle\phi_{2}\rangle|=
\frac{\sqrt{2}}{\kappa},\hspace{0.2in}\langle\phi_{3,4,..}\rangle=0, 
\end{eqnarray}
with $\phi_0$ left undetermined. In the vacuum sector supergravity is also
broken, since the gravitino acquires a mass,
\begin{equation}
m_{3/2}=\frac{1}{\kappa}e^{G/2}=\frac{e^{3/2}m^2}{3|\phi_0+\phi_0^*|^{3/2}
{\kappa}^{1/2}}.  
\end{equation}
Since we have broken a gauged $U(1)_R$ symmetry, vortex configurations will be
generated. The winding number $n$ vortex configuration is characterized by the
following behaviour of the fields :
\begin{eqnarray}
&&\phi_1(\vec{r})=e^{in\theta}f_1(r)\label{vortexfields}\\\nonumber
\\\nonumber
&&\phi_2(\vec{r})=e^{-in\theta}f_2(r)\\\nonumber
\\\nonumber
&&f_{1,2}(r) \longrightarrow
f^{0}_{1,2}r^{|n|},\hspace{0.2in}\text{as}\hspace{0.2in} 
r\rightarrow 0\\\nonumber
&&f_{1}(r) \longrightarrow \frac{1}{\kappa},\hspace{0.2in}\text{as}
\hspace{0.2in} r\rightarrow \infty\\\nonumber 
&&f_{2}(r) \longrightarrow
\frac{\sqrt{2}}{\kappa},\hspace{0.2in}\text{as}\hspace{0.2in} r\rightarrow
\infty\\\nonumber 
\end{eqnarray}
We use the function ${\cal{F}}$ to denote the generalized interactions of the
gravitino with the scalar sector of the theory (see eq. (\ref{gravitinolag})),
which gives rise to the mass term for the gravitino in the vacuum sector:
\begin{eqnarray}
&&{\cal{F}}(\phi,\phi^*)=\frac{1}{\kappa}e^{G/2}
{\left(\frac{g}{g^*}\right)}^{1/2}\label{yukawa} 
\\\nonumber
\\\nonumber
&&{\longrightarrow}\hspace{0.1in}f_0\frac{\kappa^{1/2}m^2r^{|n|}e^{in\theta}} 
{|\phi_0+\phi_0^*|^{3/2}} 
\hspace{0.2in}\text{as}\hspace{0.1in}{r{\rightarrow}0}\\\nonumber
\\\nonumber
&&{\longrightarrow}\hspace{0.1in}\frac{e^{3/2}m^2e^{in\theta}}
{3|\phi_0+\phi_0^*|^{3/2} 
{\kappa}^{1/2}} 
\hspace{0.2in}\text{as}\hspace{0.1in}{r{\rightarrow}\infty}.\\\nonumber
\end{eqnarray}

\subsubsection{CaseII: $U(1)_R$ Broken and SUGRA Unbroken}

We will now provide another simple example where the $R$-symmetry is broken
spontaneously due to the fact that one of the fields gets a vacuum expectation
value and gives rise to a string configuration. However, in this example
supergravity will remain as an exact symmetry, and thus the gravitino will
remain 
massless. We assume that the theory has 3 gauge singlets $\{S_1,S_2,S_3\}$ and
$k$ other superfields $\{\Phi_i\}$ ($i=1,2,...k$). The kinetic terms for all
the fields are taken to be canonical. The K$\ddot{\text{a}}$hler potential and
the superpotential are chosen to be:
\begin{eqnarray}
&G&(\phi,\phi^*)=\Sigma s_is_i^* \kappa^2+\Sigma\phi_i\phi
_i^* \kappa^2+\ln[\kappa^6|g(\phi_i)|^2]\\\nonumber
&g&=s_1s_2s_3
\left[m^2(\phi_1-\frac{{\sqrt{2}}\kappa^2}{3}\phi_1^2\phi_2)+\sum_{i=3}^k
\lambda_i\phi_i^2\phi_2\right]\\
&m&\sim{\cal{O}}(1/\kappa).
\end{eqnarray}
The superpotential has the same form as in the previous example except
for the overall multiplication by the singlet fields. Thus the charges of the
scalar fields $\phi_1$ and $\phi_2$ are $2$ and $-2$ repectively, while all the
remaining fields $\phi_3,\phi_4...$ have charge$=2$. The scalar potential
obtained from this particular choice of the superpotential, is

\begin{eqnarray}
V&=&e^{\kappa^2\Sigma(|\phi_i|^2+|s_i|^2)}
|g|^2\left[\sum_{i=1}^{k}\left|\kappa^2\phi_i^* 
+\frac{g^i}{g}\right|^2+
\sum_{i=1}^{3}\left|\kappa^2s_i^* 
+\frac{1}{s_i}\right|^2-3\kappa^2\right]\\\nonumber
&+&\frac{1}{2\kappa^4}e^2\left(\sum_{i=1}^{k}n_i\phi_i\left[\kappa^2\phi_i^* 
+\frac{g^i}{g}\right]\right)^2.\nonumber
\end{eqnarray}      
It is easily seen that this scalar potential is always positive, and its tree
level minimum value is at zero. A field configuration that minimizes the
potential is given by
\begin{eqnarray}
|\langle\phi_{1}\rangle|=
\frac{1}{\kappa},\hspace{0.2in}|\langle\phi_{2}\rangle|=
\frac{\sqrt{2}}{\kappa},\hspace{0.2in}\langle\phi_{3,4,..}\rangle=
\langle s_{1,2,3}\rangle=0. 
\end{eqnarray}
The anomaly cancellation conditions require
\begin{eqnarray}
TrR&=&\sum_{i=3}^kf_i-1\times3+1-3+1-21=\sum_{i=3}^kf_i-25=0\\\nonumber
TrR^3&=&\sum_{i=3}^kf_i^3+3(-1)^3+1^3+(-3)^3+1^3+3\times1^3=
\sum_{i=3}^kf_i^3-25=0,\\\nonumber
\end{eqnarray}
where $f_i=n_i-1=1$, represents the $R$-charges of the fermions, as before. But
now we need $k=27$ to cancel all the anomalies. The field configurations
characterizing the vortex are exactly the same as in eq.
(\ref{vortexfields}). However the mass term for the gravitino is exactly zero:
\begin{eqnarray}
{\cal{F}}(\phi,\phi^*)=\frac{1}{\kappa}e^{G/2}
{\left(\frac{g}{g^*}\right)}^{1/2}=0.
\end{eqnarray}

Thus we see that models can be constructed which incorporate an anomaly free
gauged R-symmetry which is spontaneously broken and that the supersymmetry can
either be broken or unbroken at the same time.

\section{Rarita-Schwinger equations in the string background}

We now turn to the question of finding solutions to the gravitino equation of
motion in the background of a vortex configuration for the $R$ gauge
field. This entails looking at the Rarita-Schwinger equation coupled to the
R-photon. This equation takes the form:

\begin{eqnarray}
{\cal{L}}&=&-i{\bar{\psi}}_{{\mu}L}\gamma^{\mu\nu\lambda}[\partial_{\nu}
-ieA_{\nu}]{\psi}_
{{\lambda}
L}-i{\bar{\psi}}_{{\mu}R}\gamma^{\mu\nu\lambda}[\partial_{\nu}+ieA_{\nu}]
{\psi}_{{\lambda}R}\label{raritalag}\\\nonumber
\\\nonumber
&+&{\cal{F}}(\phi,\phi^*){\bar{\psi}}_{{\mu}L}\gamma^{\mu\nu}{\psi}_{{\nu}R}
+{\cal{F^{*}}}(\phi^{*},\phi){\bar{\psi}}_{{\mu}R}
\gamma^{\mu\nu}{\psi}_{{\nu}L}\\\nonumber
\end{eqnarray}
We have introduced the coupling of the gravitino to the scalars through the
arbitrary complex function of the scalar fields, $\cal{F}(\phi,\phi^{*})$,
which was defined in eq. (\ref{yukawa}).  However, since the Lagrangian has a
$U(1)_R$ symmetry, $\cal{F}$ must have $R$-charge $2$. Thus the Lagrangian is
invariant under,
\begin{eqnarray}
{\psi}_{\mu}\hspace{0.1in}&{\rightarrow}&\hspace{0.1in}
\exp(-ie{\gamma}_{5}\Lambda){\psi}_{\mu}\\\nonumber
A_{\mu}\hspace{0.1in}&{\rightarrow}&\hspace{0.1in}
A_{\mu}-{\partial}_{\mu}\Lambda\\\nonumber
\cal{F}\hspace{0.1in}&{\rightarrow}&\hspace{0.1in}
\exp(-2ie\Lambda)\cal{F}.\nonumber
\end{eqnarray} 
We now suppose that the R-photon and the scalar fields take on the appropriate
form to represent a vortex solution. We then require that:

\begin{eqnarray}
&{\cal{F}}&=e^{iK\theta}{\cal{H}}(r),\label{vortexsc*}\\\nonumber
\\\nonumber
&{\cal{H}}&(r)\longrightarrow {\cal{H}}_{0}r^{|K|},\hspace{0.2in}\text{as}
\hspace{0.2in} r\rightarrow 0\\\nonumber
&{\cal{H}}&(r)\longrightarrow {\cal{H}}_{\infty},\hspace{0.2in}\text{as}
\hspace{0.2in} r\rightarrow \infty\nonumber
\end{eqnarray}
where ${\cal{H}}(r)$ is a real function, ${\cal{H}}_{0}$,
${\cal{H}}_{\infty}>0$ are constants and $K$ must be an integer to ensure that
the interaction terms are single-valued. The integer $K$ depends not only on
the vortex winding number $n$, but also on the exact form of the function
${\cal{F}}$.  If we assume that the string configuration has cylindrical
symmetry about the $z$-axis, the gauge field for the vortex solution, will only
have an azimuthal component:

\begin{eqnarray}
&\vec{A}(\vec{r})&=A(r){\hat{e}}_{\theta}.\label{gaugefld}\\\nonumber
&A(r)&{\longrightarrow}0\hspace{0.2in}\text{as}
\hspace{0.2in} r\rightarrow 0\\\nonumber
&A(r)&{\longrightarrow}\frac{n}{qr}\hspace{0.2in}\text{as}
\hspace{0.2in} r\rightarrow \infty\\\nonumber   
\end{eqnarray}
where $q$ is the charge of the scalar field that gets a VEV.  The gauge field
goes to zero at the origin and at infinity, and carries a net topological flux
$n/2q$.  From (\ref{raritalag}) we obtain the Rarita-Schwinger equations for
the spinor fields,

\begin{eqnarray}
i\gamma^{\mu\nu\lambda}[\partial_{\nu}
-ieA_{\nu}]{\psi}_{{\lambda}L}-{\cal{F}}(\phi,\phi^*)
\gamma^{\mu\nu}{\psi}_{{\nu}R}=0,
\label{raritaeq}\\\nonumber
\\\nonumber
i\gamma^{\mu\nu\lambda}[\partial_{\nu}+ieA_{\nu}]
{\psi}_{{\lambda}R}-{\cal{F^{*}}}(\phi^{*},\phi)
\gamma^{\mu\nu}{\psi}_{{\nu}L}=0. 
\\\nonumber
\end{eqnarray}
In the vacuum sector, the fermions get a mass ${\cal{H}}_{\infty}$, while in
the string core, they are massless. We will solve eq.(\ref{raritaeq}) with a
background gauge field given by eq.(\ref{gaugefld}) in the next section.

\subsection{Solutions to the Rarita-Schwinger equations}

{\em Case I (SUGRA unbroken)}:
\\

In this case, the gravitino obeys the massless Rarita-Schwinger equations,
\begin{eqnarray}
i\gamma^{\mu\nu\lambda}[\partial_{\nu}
-ieA_{\nu}]{\psi}_{{\lambda}L}=0,
\label{massless}\\\nonumber
\\\nonumber
i\gamma^{\mu\nu\lambda}[\partial_{\nu}+ieA_{\nu}]
{\psi}_{{\lambda}R}=0.
\nonumber
\end{eqnarray} 
These equations can be rewritten in terms of the gauge covariant derivatives as
\begin{eqnarray}
&&\gamma^{\mu\nu\lambda}D_{\nu}{\psi}_{{\lambda}L}
\label{masslesseq}\\\nonumber
=&&\gamma^{\nu}D_{\nu}{\psi}^{\mu}_{L}+\gamma^{\mu}\gamma^{\nu\lambda}
D_{\nu}{\psi}_{{\lambda}L} -D^{\mu}(\gamma\cdot{\psi}_{L})=0
\end{eqnarray}
One can get the right-handed field equations by replacing
the covariant derivatives $D_i$ with $D^*_i$.
Contracting this equation with $\gamma_{\mu}$ yields
\begin{eqnarray}
&&\gamma^{\nu\lambda}D_{\nu}{\psi}_{{\lambda}L}=0\\\nonumber
&&\gamma^{\nu}D_{\nu}{\psi}^{\mu}_{L}=D^{\mu}(\gamma\cdot{\psi}_{L}).
\end{eqnarray}
Now, acting on eq. (\ref{masslesseq}) with the operator $D_{\mu}$, we get,
\begin{eqnarray}
\gamma^{\mu\nu\lambda}D_{\mu}D_{\nu}{\psi}_{{\lambda}L}
=-\frac{ie}{2}\gamma^{\mu\nu\lambda}F_{\mu\nu}{\psi}_{{\lambda}L}=0.
\end{eqnarray}
But the vector potential has only an azimuthal component which means that the
only non-zero components of $F_{\mu\nu}$ are $F_{12}$ and $F_{21}$. Therefore,
we find,
\begin{equation}
\gamma^0{\psi}_{{0}L,R}+\gamma^3{\psi}_{{3}L,R}=0.
\end{equation}
If we impose the additional condition that $\gamma\cdot{\psi}_{L,R}=0$ i.e.
$\gamma^1{\psi}_{{1}L,R}+\gamma^2{\psi}_{{2}L,R}=0$, we find that each of the
$4$ components of the spinor satisfies the Dirac equation
\begin{eqnarray}
\gamma^\nu D_\nu\psi^{\mu}_{L,R}=0.
\end{eqnarray}
Now we can exploit the symmetry of the problem and assume that the solutions
are separable i.e. $\psi_{{\mu}L,R}=\beta(z,t)\tilde{\psi}_{{\mu}L,R}(x,y)$.
One can easily see that the spinors must then satisfy the transverse Dirac
equation,
\begin{eqnarray}
(\gamma^1D_1+\gamma^2D_2)\psi_{{\mu}L}=0,\label{transdirac}
\\\nonumber
\\\nonumber
(\gamma^1D_1^*+\gamma^2D_2^*)\psi_{{\mu}R}=0.
\end{eqnarray}
In addition, the left-handed fields have to satisfy
\begin{equation}
(\partial_0+\sigma_3\partial_3)\psi_{{\mu}L}=0,
\end{equation}
while the right-handed fields satisfy
\begin{equation}
(\partial_0-\sigma_3\partial_3)\psi_{{\mu}R}=0.
\end{equation}

We now have the option of choosing the fields to be eigenstates of $\sigma_3$.
A left-handed spin-up (down) state ends up being a right (left)-mover, while a
right-handed spin-up (down) state is a left (right)-mover. We will focus only
on the left-handed spin-up fields -- the other cases can be handled in a
similar fashion. Thus we take:

\begin{eqnarray}
\psi_{{\mu}L}={\alpha(z-t)}\left(\begin{array}{c}
u(r,\theta)
\\\nonumber 
0\end{array}\right).\\\nonumber
\end{eqnarray}
Since the spinors are zero modes of the transverse Dirac operator
(\ref{transdirac}), the function $u(r,\theta)$ is a solution to the
differential equation:
\begin{eqnarray}
[\partial_{r}+eA(r)+\frac{i}{r}\partial_{\theta}]u(r,\theta)=0,
\label{tdirac}
\end{eqnarray}
which implies that
\begin{eqnarray}
u(r,\theta)&=&{\sum}_{m}a_mr^me^{im\theta}\exp(-e\int^rA(r)dr),\label{sol1}
\\\nonumber
m&=&0,1,2....[\frac{ne}{q}-1].\nonumber
\end{eqnarray}
where [N] denotes the greatest integer smaller than N. The right-moving
solutions are therefore given by
\begin{eqnarray}
\psi_{1L}&=&-i\psi_{2L}=-i\alpha(z-t)\left(\begin{array}{c}
u(r,\theta)\\\nonumber
0\end{array}\right),
\\\label{sol}
\\\nonumber
\psi_{0L}&=&-\psi_{3L}=-\beta(z-t)\left(\begin{array}{c}
u_1(r,\theta)\\\nonumber
0\end{array}\right),\nonumber
\\\nonumber
\\\nonumber
\psi_{1R}&=&i\psi_{2R}=i\alpha_1(z-t)\left(\begin{array}{c}
0\\\nonumber
v(r,\theta)\end{array}\right),\nonumber
\\\nonumber\\\nonumber
\psi_{0R}&=&-\psi_{3R}=-\beta_1(z-t)\left(\begin{array}{c}
0\\\nonumber
v_1(r,\theta)\end{array}\right).\nonumber
\\\nonumber\\\nonumber
\end{eqnarray}
Here $u$, $u_1$ are solutions to (\ref{tdirac}), while $v$, $v_1$ are solutions
to
\begin{equation}
[\partial_r+eA(r)-\frac{i}{r}\partial_{\theta}]v=0 \nonumber
\end{equation}

It is not difficult to verify that (\ref{sol}) satisfy the covariant condition
$D\cdot \psi=0$.  The solutions (\ref{sol1}) are normalizable provided
${ne}\slash {q}>1$ since asymptotically far from the string,
$A(r)={n}\slash {qr}$ and $u{\sim}r^{-ne/q}$.

The left-handed fermionic current along the z-axis is given by
$=-e{\bar{\psi}}_{{\mu}L}\gamma^{{\mu}3\lambda}\psi_{{\lambda}L}$ and for the
spin-up solutions (\ref{sol}) is given by $e|{\alpha}(z-t)| ^2
|u(r,\theta)|^2$. In the absence of a clear proof, we speculate that these
non-zero currents could lead to superconducting/persistent $R$-currents in the
string.

The calculations so far, have closely parallelled that of reference
\cite{jackiw2}, where the zero modes of the massless Dirac equation were
studied in the presence of a magnetic field in two spatial dimensions. We will
now look at the case where supersymmetry is broken and the gravitino becomes
massive.  \\ \\ {\em Case II (SUGRA broken)}: \\

In this scenario, as indicated in eqns.(\ref{raritaeq}), the gravitino acquires
a mass away from the string core. Consequently, the equations for the
left-handed and right-handed fermions are coupled and we have a much more
complicated situation than the SUSY unbroken case. However, based on the
previous calculation we can hazard a guess as to the form of the solutions --
in particular, we expect that there will be solutions which are eigenstates of
$\sigma_3$ in the presence of the $R$ magnetic field.  We thus propose the
following ansatz:

\begin{eqnarray}
\psi_{{\mu}L}=\left(\begin{array}{c}U_{\mu}\\
0\end{array}\right)\\\nonumber
\\\nonumber 
\psi_{{\mu}R}=\left(\begin{array}{c}0\\
V_{\mu}\end{array}\right)
\end{eqnarray}
Writing out the field equations in terms of the spinor components we get,
\begin{eqnarray}
(i&)&\hspace{.5in}D_1U_2-D_2U_1-2{\cal{F}}(V_1-iV_2)=0\label{leftcomp}
\\\nonumber
\\\nonumber
(ii&)&\hspace{.5in}(D_2-iD_1)U_3-\partial_3(U_2-iU_1)+2{\cal{F}}V_3=0
\\\nonumber
\\\nonumber
(iii&)&\hspace{.5in}\partial_3U_0-\partial_0U_3-2{\cal{F}}V_2=0
\\\nonumber
\\\nonumber
(iv&)&\hspace{.5in}(\partial_0+\partial_3)U_2-D_2(U_0+U_3)-2{\cal{F}}
(V_0+V_3)=0 
\\\nonumber
\\\nonumber
(v&)&\hspace{.5in}\partial_3U_0-\partial_0U_3-2i{\cal{F}}V_1=0
\\\nonumber
\\\nonumber
(vi&)&\hspace{.5in}iD_1(U_0+U_3)-i(\partial_0+\partial_3)U_1-2{\cal{F}}
(V_3+V_0)=0  
\\\nonumber
\\\nonumber
(vii&)&\hspace{.5in}\partial_0(U_2-iU_1)-(D_2-iD_1)U_0-2{\cal{F}}V_0=0.
\\\nonumber
\end{eqnarray}

\begin{eqnarray}
(a&)&\hspace{.5in}D_1^*V_2-D_2^*V_1-2{\cal{F}}^*(U_1+iU_2)=0\label{rightcomp}
\\\nonumber
\\\nonumber
(b&)&\hspace{.5in}(D_2^*+iD_1^*)V_3-\partial_3(V_2+iV_1)+2{\cal{F}}^*U_3=0
\\\nonumber
\\\nonumber
(c&)&\hspace{.5in}\partial_3V_0-\partial_0V_3-2{\cal{F}}^*U_2=0
\\\nonumber
\\\nonumber
(d&)&\hspace{.5in}(\partial_0+\partial_3)V_2-D_2^*(V_0+V_3)-2{\cal{F}}^*
(U_0+U_3)=0 
\\\nonumber
\\\nonumber
(e&)&\hspace{.5in}\partial_3V_0-\partial_0V_3+2i{\cal{F}}^*U_1=0
\\\nonumber
\\\nonumber
(f&)&\hspace{.5in}D_1^*(V_0+V_3)-(\partial_0+\partial_3)V_1-2i
{\cal{F}}^*(U_3+U_0)=0   
\\\nonumber
\\\nonumber
(g&)&\hspace{.5in}\partial_0(V_2+iV_1)-(D_2^*+iD_1^*)V_0-2{\cal{F}}^*
U_0=0 
\end{eqnarray}
The equations (\ref{leftcomp}) $(iii)$, (\ref{leftcomp}) $(v)$,
(\ref{rightcomp}) $(c)$ and (\ref{rightcomp}) $(e)$ imply that $U_1=iU_2$ and
$V_1=-iV_2$. Using these relations and subtracting eq.$(vii)$ from eq.$(ii)$
we get,
\begin{equation}
(D_2-iD_1)(U_3+U_0)-2(\partial_3+\partial_0)U_2+2{\cal{F}}(V_0+V_3)=0.
\label{fst}
\end{equation}
On the other hand eqns.$(iv)$, $(vi)$ yield 
\begin{equation}
(D_2-iD_1)(U_3+U_0)-2(\partial_3+\partial_0)U_2+4{\cal{F}}(V_0+V_3)=0.
\end{equation}
We therefore find that $V_0=-V_3$. A similar manipulation of
eqns. (\ref{rightcomp}) $(b, d, f, g)$ shows that $U_0=-U_3$. This, then
implies that
\begin{eqnarray}
(\partial_0+\partial_3)U_2=0,\\\nonumber
(\partial_0+\partial_3)V_2=0,
\end{eqnarray}
i.e. $U_2=U_2(x,y,z-t)$ and $V_2=V_2(x,y,z-t)$.
Let us now define the light-cone variables $X_+=(z+t)/2$ and $X_-=(z-t)/2$.
In terms of these variables eq. (\ref{leftcomp}) $(iii)$ can be written as 
\begin{eqnarray}
\partial_+U_0={\cal{F}}V_2(X_-).
\end{eqnarray}
The solution to this equation is then
\begin{eqnarray}
U_0(X_+,X_-)={\cal{F}}V_2(X_-)X_++g(X_-)
\end{eqnarray}
where $g$ is an arbitrary function of $X_-$. But this form of $U_0$ is in
general unbounded in $z$ and $t$ because of the linear dependence on $X_+$,
unless $V_2$ vanishes identically. Using a similar argument, one can show that
$U_2$ must also be zero. These results can be shown to hold for the spin-down
fields too, except that they happen to be left-movers. Thus the solutions can
be characterised by the following set of conditions:

\begin{eqnarray}
{\psi}_{1L,R}&=&{\psi}_{2L,R}=0\label{ansatz1},
\\\nonumber
\psi_{0L}&=&-\sigma_3\psi_{3L}=\pm\psi_{3L}\\\nonumber
&=&\alpha(z{\pm}t){\tilde{\psi}}_{L}(x,y)\\\nonumber \text{and}\\\nonumber
\psi_{0R}&=&\sigma_3\psi_{3R}=\pm\psi_{3R}\\\nonumber
&=&\alpha(z{\pm}t){\tilde{\psi}}_{R}(x,y).\\\nonumber \\\nonumber
i(\sigma_1D_1+\sigma_2D_2)&{\tilde{\psi}}_{L}&(x,y)-2{\cal{F}}{\tilde{\psi}}_
{R}(x,y)=0 \\\nonumber i(\sigma_1D_1^*+\sigma_2D_2^*)&{\tilde{\psi}}_{R}&(x,y)+
2{\cal{F}}^{*}{\tilde{\psi}}_{L}(x,y)=0.\\\nonumber
\end{eqnarray}
Let us first try to obtain the right-moving solutions explicitly, i.e.
\begin{eqnarray}
\psi_{0L}&=-&\psi_{3L}=\alpha(z-t)\left(\begin{array}{c}u(x,y)\\0\end{array}
\right)\label{ansatz}
\\\nonumber
\text{and}\\\nonumber
\psi_{0R}&=-&\psi_{3R}=\alpha(z-t)\left(\begin{array}{c}0\\-iv(x,y)\end{array}
\right).
\\\nonumber
\end{eqnarray}
This would describe a left-handed spin-up Weyl spinor, and a right-handed
spin-down Weyl spinor moving along the positive $z$ axis at the speed of light.
Using (\ref{leftcomp}), (\ref{rightcomp}) and (\ref{ansatz1}), we can obtain
the equations for the complex functions $u$ and $v$ in polar-coordinates
defined on the $x$-$y$ plane:
\begin{eqnarray}
&\partial_ru&+\frac{i}{r}\partial_{\theta}u+eA(r)u+2e^{i(K-1)\theta}
{\cal{H}}v=0\label{a}\\
&\partial_rv&-\frac{i}{r}\partial_{\theta}v+eA(r)v+2e^{-i(K-1)\theta}
{\cal{H}}u=0
\label{b}
\end{eqnarray}
Adding (\ref{a}) and the complex conjugate of (\ref{b}) we obtain,
\begin{equation}
\partial_r\delta+\frac{i}{r}\partial_{\theta}\delta+eA(r)\delta
+2e^{i(K-1)\theta}
 {\cal{H}}\delta^*=0\label{c}
\end{equation}
where, $\delta=(u+v^*)$.  On the other hand, subtracting the complex conjugate
of (\ref{b}) from (\ref{a}) gives,

\begin{equation}
\partial_r\eta+\frac{i}{r}\partial_{\theta}\eta+eA(r)\eta
-2e^{i(K-1)\theta}
 {\cal{H}}\eta^*=0\label{d}
\end{equation}
where $\eta=u-v^*$. We can now follow the analysis of \cite{jackiw} to obtain
the solutions to (\ref{c} and \ref{d}). We write $\eta$ and $\delta$ as
\begin{eqnarray}
\delta&=&e^{-e{\int}A(r)dr}(\delta_1e^{im\theta}+\delta_2e^{i(K-1-m)\theta})
\label{reduction}\\\nonumber
\eta&=&e^{-e{\int}A(r)dr}(\eta_1e^{im\theta}+\eta_2e^{i(K-1-m)\theta})
\\\nonumber
\end{eqnarray} 
where $m$ is an integer $\neq\frac{K-1}{2}$. We notice that (\ref{d}) can be
obtained from (\ref{c}) by the replacement,
${\cal{H}}\rightarrow-{\cal{H}}$. It therefore, suffices to concentrate on the
solutions of (\ref{c}). Without any loss of generality, we assume that
asymptotically away from the string,
${\cal{H}}\rightarrow{\cal{H}}_{\infty}>0$.

Since the differential equations involve complex conjugation, they are
non-linear and hence need to be linearized. This can be done by writing them in
terms of the real and imaginary parts of the $\delta$'s, namely $\delta_{1,2r}$
and $\delta_{1,2i}$ thus reducing (\ref{c}) to the pair of equations:
\begin{eqnarray}
&\delta^{\prime}_{1r(i)}&-\frac{m}{r}\delta_{1r(i)}+(-)2{\cal{H}}
\delta_{2r(i)}=0\\\nonumber
&\delta^{\prime}_{2r(i)}&-\frac{K-1-m}{r}\delta_{2r(i)}+(-)2{\cal{H}}
\delta_{1r(i)}=0\\\nonumber
\end{eqnarray}
where the prime denotes a radial derivative. The corresponding
equations for the $\eta$ variables are
\begin{eqnarray}
&\eta^{\prime}_{1r(i)}&-\frac{m}{r}\eta_{1r(i)}-(+)2{\cal{H}}
\eta_{2r(i)}=0\\\nonumber
&\eta^{\prime}_{2r(i)}&-\frac{K-1-m}{r}\eta_{2r(i)}-(+)2{\cal{H}}
\eta_{1r(i)}=0.\\\nonumber
\end{eqnarray} 
Following the arguments in \cite{jackiw} we define a pair of real normalizable
functions $(u_m,v_m)$ which satisfy,
\begin{eqnarray}  
&u^{\prime}_{m}&-\frac{m}{r}u_{m}+2{\cal{H}}
v_{m}=0\label{mutilatedeq.}\\\nonumber
&v^{\prime}_{m}&-\frac{K-1-m}{r}v_{m}+2{\cal{H}}
u_{m}=0.\\\nonumber
\end{eqnarray}
The general solutions for $K>0$ are then given by:
\begin{eqnarray}
\delta_{1}&=&a_{m}u_{m},\hspace{0.2in}\delta_{2}=a_m^*v_m
\eta_{1}=b_{n}u_{n},\hspace{0.2in}\eta_{2}=-b_n^*v_n\\\nonumber
\text{and}\\\nonumber
\delta&=&e^{-e{\int}A(r)dr}[a_{m}u_{m}(r)e^{im\theta}+a_m^*v_m(r)
e^{i(K-1-m)\theta}]\\\nonumber
\\\nonumber
\eta&=&e^{-e{\int}A(r)dr}[b_{n}u_{n}(r)e^{in\theta}-b_n^*v_n(r)
e^{i(K-1-n)\theta}]\\\nonumber
\\\nonumber 
K&-&1{\geq}m,n{\geq}0.
\end{eqnarray}
The $a$'s and $b$'s are constants of integration.  The functions $u_m,v_m$ must
be well behaved both at infinity and at the origin; this forces $\eta$ and
$\delta$ to go to zero, whenever $K<0$. The final form of the right-moving
solutions thus obtained is:
\begin{eqnarray}
&\psi_{0L}&=-\psi_{3L}=\frac{\alpha(z-t)}{2}\left(\begin{array}{c}
e^{-e{\int}A(r)dr}[a_{m}u_{m}(r)e^{im\theta}+a_m^*v_m(r)
e^{i(K-1-m)\theta}\\\nonumber
\\\nonumber+b_{n}u_{n}(r)e^{in\theta}-b_n^*v_n(r)
e^{i(K-1-n)\theta}]\\\nonumber\\0\end{array}\right)\label{solutions}\\\nonumber
&\psi_{1L}&=\psi_{2L}=0\\
\\\nonumber
\text{and}\\\nonumber
&\psi_{0R}&=-\psi_{3R}=\frac{\alpha(z-t)}{2}\left(\begin{array}{c}
0\\\\\nonumber
-ie^{-e{\int}A(r)dr}[a_{m}^*u_{m}(r)e^{-im\theta}+a_mv_m(r)
e^{-i(K-1-m)\theta}\\\nonumber
\\\nonumber-b_{n}^*u_{n}(r)e^{-in\theta}+b_nv_n(r)
e^{-i(K-1-n)\theta}]\end{array}\right)\\\nonumber
&\psi_{1R}&=\psi_{2R}=0\\\nonumber
\end{eqnarray}
The case $m,n=\frac{K-1}{2}$ has to be treated separately, because the ansatz
(\ref{reduction}), now involves only one function. Obviously, this can happen
only when $K$ is odd. Modifying the ansatz, we obtain,

\begin{eqnarray}
\delta&=&e^{-e{\int}A(r)dr}\delta_1e^{i\frac{(K-1)}{2}\theta}
\\\nonumber
\eta&=&e^{-e{\int}A(r)dr}\eta_1e^{i\frac{(K-1)}{2}\theta}
\\\nonumber
\end{eqnarray}
The solutions are easy to obtain and fit into the form shown in
(\ref{solutions}). In particular, we see that $u_{K/2-1/2}(r)=v_{K/2-1/2}(r)=
r^{K/2-1/2}\exp[-2{\int}{\cal{H}}(r)dr]$. Therefore, we have seen that the
right-moving solutions given by (\ref{ansatz}) exist only when $K>0$.

What about left-movers? Going through an analysis similar to the above we see
that left-movers are allowed only when $K<0$. Defining $K^{\prime}=-K>0$ we
obtain the following as the solutions,
\begin{eqnarray}
&\psi_{0L}&=\psi_{3L}=\frac{\alpha(z+t)}{2}\left(\begin{array}{c}
0\\\\\nonumber
e^{e{\int}A(r)dr}[c_{m}u_{m}(r)e^{-im\theta}+c_m^*v_m(r)
e^{-i(K^{\prime}-1-m)\theta}\\\nonumber
\\\nonumber+d_{n}u_{n}(r)e^{-in\theta}-d_n^*v_n(r)
e^{-i(K^{\prime}-1-n)\theta}]\end{array}\right)
\\\nonumber
&\psi_{1L}&=\psi_{2L}=0\\
\\\nonumber
\text{and}\\\nonumber
&\psi_{0R}&=\psi_{3R}=\frac{\alpha(z+t)}{2}\left(\begin{array}{c}
-ie^{e{\int}A(r)dr}[c_{m}^*u_{m}(r)e^{im\theta}+c_mv_m(r)
e^{i(K^{\prime}-1-m)\theta}\\\nonumber
\\\nonumber-d_{n}^*u_{n}(r)e^{in\theta}+d_nv_n(r)
e^{i(K^{\prime}-1-n)\theta}]\\\nonumber\\0\end{array}\right)\\\nonumber
&\psi_{1R}&=\psi_{2R}=0\\\nonumber
\\\nonumber
&K^{\prime}&-1{\geq}m,n{\geq}0.
\end{eqnarray}
Another way to obtain left-movers is to have fermions with $R$-charge $-e$ and
couple them to ${\cal{F}}^*$.  One can easily check that these solutions
satisfy the conditions $D\cdot \psi=0$ and $\gamma \cdot \psi=0$. However, the
charge density and current density for these solutions, given by
$-e{\bar{\psi}}_{{\mu}L} \gamma^{{\mu}3\lambda}\psi_{{\lambda}L}$ happens to
be identically zero. Thus they cannot give rise to superconducting currents of
any kind.

There are two points to note about the solutions that we have obtained. First,
they are simply the zero modes of the transverse Dirac operator in a vortex
background, and the properties of these zero modes are well-known, both from
index theorems and from explicit solutions \cite{jackiw},
\cite{weinberg}. Second, as one sees in the case of spin-$1/2$ fermions, the
solutions (\ref{solutions}) are localized along the string. The functions are
exponentially suppressed as one moves away from the string, being proportional
to $\exp[-{\cal{H}}_0r]$ asymptotically. Since ${\cal{H}}_0^{-1}$ will
typically be of the order of the Compton wavelength of the fermion, we have a
fermionic bound state wavefunction that has practically all its support
concentrated at the core of the string.
   
\section{Conclusions and further questions}

In this article we have studied some of the solutions to the spin-$3/2$
equation of motion in a cosmic string background. Our motivation was to look
for physically interesting situations ( for e.g. superconducting currents) in
the context of supergravity theories with a gauged $R$-symmetry, where the
gravitino ends up having a non-zero $R$-charge.  We have also provided an
explicit construction of toy models of supergravity with gauged
$R$-symmetry. The particle contents of the theories were chosen to cancel the
$U(1)_R$ and mixed gravitational anomalies, thus leading to models where the
symmetries are implemented in a consistent fashion. In the first example, both
supergravity and $R$-symmetry were broken, while the second model had vacua
where the supergravity was unbroken, leading to string configurations which
were obtained explicitly. Such configurations could also arise in more
realistic models, such as the extensions to MSSM considered in \cite{dreiner}.

We obtained solutions to the classical field equations (Rarita-Schwinger
equations) for the gravitino in the vortex background, in both the supergravity
unbroken and broken phases. In both cases, the solutions possessed similarities
with previously obtained solutions to the Dirac equation in vortex backgrounds,
simply because they all turned out to be zero-modes of the transverse Dirac
operator. In the SUGRA unbroken phase, the field configurations carried
non-zero 
fermionic-currents along the string, pointing to the possibility of having
superconducting states in the string. In the supergravity broken phase,
however, we found solutions that did not carry any fermionic current or charge
density, precluding the possibility of having persistent/superconducting
$R$-currents, at least for the class of solutions found by us.

Our studies point to further interesting questions: 

\begin{enumerate}
\item{Is it possible to get solutions other than the ones already obtained,
that will give rise to persistent currents? A closely related issue that would
have to be understood is, how one would go about proving the existence of these
superconducting currents. The answers are well-known for the spin-$1/2$ case,
but the spin-$3/2$ case is intrinsically different and is yet to be
understood.}

\item{There are other mathematically stimulating problems, namely index
theorems for the Rarita-Schwinger equation in a string background.}
\end{enumerate}

We have concerned ourselves mostly with the interesting mathematical question
of whether given gravitinos with non-zero R charges, they can give rise to
trapped zero modes on R strings. However, there may be constraints on the
existence of R strings in the universe depending on when they form. If the
R-symmetry has to be broken at a scale greater than $10^{16}$ GeV, as may be
required from considerations of naturalness\cite{dreiner,dzf2}, the energy
density in strings may be to large to be accomodated by observations and the
strings may have to be inflated away\cite{vilshell}. If use is to be made of
gravitino zero modes, the symmetry breaking scale will have to be reduced {\em
without} fine tuning the theory. It remains to be seen whether such models
exist.

\section{acknowledgements}
R.H. would like to thank H. Dreiner for stimulating conversations.
The authors would also like to thank D. Boyanovsky and H. J. de Vega for
illuminating discussions, and 
useful references. 
This work was supported in part by DOE Grant $\#$ DE-FG02-91-ER40682. 
  
\end{document}